\documentclass[onecolumn]{revtex4}
\usepackage{graphicx}
\usepackage{amssymb}

\def\al{\alpha}

\def\Ga{\Gamma}

\def\si{\sigma}

\def\La{\Lambda}

\def\bx{{\bf{x}}}

\def\E{{\rm E}_{10}}

\def\rS{{\rm\bf S}}

\def\cO{{\mathcal O}}

\def\11{{\mathbb 1}}

\def\re{{\rm e}}

\def\rd{{\rm d}}

\def\E{{\bf 1}}

\def\eV{\,\rm eV}

\def\GeV{\,\rm GeV}

\def\MPL{M_{\rm Pl}}

\def\beq{\begin{equation}}
\def\eeq{\end{equation}}
\def\bea{\begin{eqnarray}}
\def\eea{\end{eqnarray}}

\def\Ra{\Rightarrow}

\begin{document}
\title{Superheavy Gravitinos and Ultra-High Energy Cosmic Rays}
\author{Krzysztof A. Meissner$^1$ and Hermann Nicolai$^2$}
\address{
$^1$Faculty of Physics,
University of Warsaw\\
Pasteura 5, 02-093 Warsaw, Poland\\
$^2$Max-Planck-Institut f\"ur Gravitationsphysik
(Albert-Einstein-Institut)\\
M\"uhlenberg 1, D-14476 Potsdam, Germany
}

\vspace{3mm}

\begin{abstract} 
\centerline{\bf Dedicated to the memory of Murray Gell-Mann}
\vspace{3mm}
\noindent
We argue that the superheavy gravitinos that we had previously proposed 
as candidates for Dark Matter can offer a possible explanation for 
the ultra-high energy cosmic ray (UHECR) events observed at the Pierre Auger 
Observatory, via gravitino anti-gravitino annihilation in the `skin' of neutron stars. 
The large mass and strong interactions of these particles, together with their stability 
against decays into standard matter are essential for the proposed explanation 
to work. In particular, it ensues that UHECR events can be understood to originate from 
neutron stars inside a GKZ horizon of $\sim$ 50 Mpc. The composition of neutron 
stars near their surface could play a crucial role in explaining the presence of heavy ions 
in these events. If confirmed, this new mechanism can be taken as evidence for the 
fundamental ansatz towards unification on which it is based.
\end{abstract}
\maketitle

\vspace{5mm}


To date there does not appear to exist a fully satisfactory 
explanation in terms of known physics for the ultra-high energy cosmic ray (UHECR) 
events observed over many years at the Pierre Auger Observatory 
\cite{AUGER,UHE1,UHE2}, which reach maximum energies of $E \gtrsim 10^{19} \eV$ saturating the GKZ bound \cite{GKZ}. A particularly puzzling feature is the occurrence of
not just protons, but heavier ions (C, N, O, or heavier) which appear to dominate 
towards the very highest energies \cite{AUGER}. The absence of a compelling scenario
may indicate the need for new physics  beyond the Standard Model (SM). In this 
article we present a possible new explanation very different from previous proposals. 
This is based on our previous work \cite{MN0,MN1} where we have raised 
the possibility that  dark matter (DM) could consist at least in part of an extremely 
dilute gas of very massive stable gravitinos, which are furthermore fractionally charged 
and strongly interacting. As we will argue here these can in principle furnish a
fairly simple explanation for the most energetic cosmic ray energy events, 
both qualitatively and quantitatively.

Possible acceleration mechanisms relying on more conventional physics, 
such as Fermi acceleration of known particles by shock waves, have been amply 
discussed in the literature \cite{UHE2}, but so far no clear picture has emerged.
A more exotic possibility is to invoke the Penrose process of extracting energy 
from a rapidly rotating Kerr black hole \cite{MTW}. Unfortunately, such an explanation 
runs into difficulties with Thorne's theorem \cite{T}, according to which rapidly spinning 
black holes obey  $a/M \lesssim \beta_{max} = 0.998$ (where $a$ is the usual rotation
parameter of the Kerr solution). It seems reasonable to assume that this bound also 
sets an upper limit for the velocity that a proton (or any other elementary particle) can 
acquire in such a process, but then the maximum attainable energy falls far short 
of the required $E \gtrsim 10^{19} \eV$ (corresponding to $1-\beta \sim 10^{-20}$). 
Another possible explanation could be via annihiliation of GUT-like objects,
such as GUT mass magnetic monopoles. However, even assuming these do exist,
it is not clear whether and how they could accumulate in sufficient amounts to explain the 
observed event rates. The same objection applies to other hypothetical GUT-like 
objects like leptoquarks or heavy gauge bosons, as these would most likely have 
decayed already long ago.  
In conclusion, there appears to be no compelling mechanism, neither from relativistic 
astrophysics nor from SM physics or widely discussed ``Beyond the SM" scenarios, that 
could plausibly explain the acceleration of known (or suspected new) particles 
to the required energies, nor account for the observed abundance of heavy ions.

The new explanation proposed here is entirely different, being based
on a more fundamental ansatz \cite{MN0,MN1}. That work
was originally motivated by an attempt to explain the fermion 
content of the SM, with three generations of 16 quarks and leptons each, from the 
spin-$\frac12$ fermion content of maximal $N\!=\!8$ supergravity, following a proposal 
orginally due to Gell-Mann \cite{GM,NW}. This proposal was further developed in 
\cite{MN2,KN,MN0} in order to fully account for the 
SU(3)$_c \times$ SU(2)$_w \times$ U(1)$_{em}$ assignments of the SM fermions, 
by exploiting properties of the maximal compact subgroup (`R symmetry') $K(E_{10})$ 
of the conjectured maximal duality symmetry $E_{10}$. Just like $N\!=\!8$ supersymmetry, 
this group theoretical framework entails the existence of eight massive gravitinos 
in addition to the 48 fundamental spin-$\frac12$ fermions. The present proposal is 
thus not simply based on {\em ad hoc} postulates, but part of a wider framework 
for unification with emergent space-time \cite{DHN};  however, (Planck scale) supersymmetry 
is here not necessarily realized as a {\em bona fide}  symmetry in the 
framework of space-time based quantum field theory. Although
it so far relies solely on group theoretic considerations (whereas 
a proper dynamical description would require a much better  understanding of 
the infinite-dimensional duality symmetries underlying it), one can nevertheless
derive some interesting consequences from this kinematic framework 
even without detailed knowledge of the dynamics.

Before continuing let us add a word of caution. Our main goal here is to point out a new 
possible mechanism for the generation of UHECRs, and not to present precise
estimates for event rates (something other proposed schemes also cannot do). 
Although we do present some numerical estimates these are by no means 
meant to be definite predictions; rather the aim is to show that, with some 
reasonable assumptions, one can arrive at event rates that are not too far from 
the ones observed. There are many uncertainties, both known and unknown,
about the actual physics, for instance concerning the composition and density 
profiles of neutron stars, unknown astrophysical input (such as the local distribution 
of DM), various unresolved issues of strong interaction dynamics, or extrapolations of 
known formulas (such as (\ref{mult}) below) far beyond their tested domains of validity. 
Any change in these numbers could significantly alter the final outcome. so the present 
exercise should be rather viewed as a ``proof of principle". For this reason we will
usually  neglect factors of $\cO(1)$ in the calculations below, as with current knowledge 
we anyhow cannot pretend to a higher level of precision.

The eight massive gravitinos are characterized by the following properties. 
From the group theoretic analysis given in \cite{MN2,KN} it follows that they transform as 
\beq\label{GravCharges}
\left({\bf 3}\,,\,\frac13\right) \oplus \left(\bar{\bf 3}\,,\,-\frac13\right)
\oplus \left({\bf 1}\,,\,\frac23\right) \oplus \left({\bf 1}\,,\, -\frac23\right)
\eeq 
under SU(3)$_c \,\times\,$U(1)$_{em}$. These assignments follow from 
an SU(3)$\,\times\,$U(1) decomposition of the $N\!=\!8$ supergravity gravitinos,
{\em except} for the shift of the U(1) charges that was originally introduced in \cite{GM} 
to make the electric charge assignments of the spin-$\frac12$ fermions agree with 
those of the quarks and leptons. As shown in \cite{MN0,MN2,KN}, it is this latter shift which 
requires enlarging the SU(8) R symmetry of $N\!=\!8$ supergravity to $K(E_{10})$, 
and which takes the construction {\em beyond} $N\!=\!8$ supergravity.
Hence, unlike DM candidates usually considered (such as axions or WIMPs),
these particles {\em do} participate in strong and electromagnetic interactions, with 
coupling strengths of order $\cO(1)$. All gravitinos are assumed to be supermassive 
with masses not too far from the (reduced) Planck mass 
$\MPL \sim 2 \cdot10^{18} \GeV/c^2 \sim 4\cdot 10^{-9} {\rm kg}$ (in a supersymmetric 
context this would correspond to Planck scale breaking of supersymmetry).
Finally, the charge assignments (\ref{GravCharges}) ensure that, despite their strong 
and electromagnetic interactions with ordinary matter, the superheavy gravitinos 
are stable because there are no fractionally charged 
final states in the SM into which they could decay in a way compatible with 
SU(3)$_c\,\times\,$U(1)$_{em}$ and (\ref{GravCharges}). Hence the only process 
that can lead to their disappearance is mutual annihilation, and this will be the crucial 
effect considered here. 

However, there is an essential difference between the two kinds of gravitinos in 
(\ref{GravCharges}). While the color singlet gravitinos would be pointlike objects
(of size $\sim \MPL^{-1}$) the color triplet  gravitinos are expected to form color singlet 
bound states with quarks, and thus very complicated dynamical objects that come with a
gluonic cloud, which would effectively enhance their size from $\MPL^{-1}$ 
to $1/\La_{QCD}$. We also note that because of their large mass these objects can be 
regarded as non-relativistic in all circumstances considered in this paper, 
whence intuition based on a parton picture does not apply, just as it does not apply to 
low energy protons -- in particular, it makes no sense to attribute the mass of the 
bound state to its constituents in any particular way. There is (so far at least) basically nothing 
known about the dynamics of such putative superheavy strongly interacting  objects, 
hence we will have to argue by analogy with known strong interaction physics,
and to work with extrapolations of known and tested formulas to explore how the strongly 
interacting gravitinos can contribute to the annihiliation processes producing UHECR events.

Both the large gravitino mass and the amount of accumulated mass of these particles are 
necessary to understand the large energies and the rates of the observed UHECR events, 
as we shall now explain. In addition we need to make one important further assumption 
concerning the {\em local} distribution of DM in stellar systems. The average density 
of DM within a typical galaxy is commonly given as
$\rho_{DM} \,\sim\,  0.3 \times 10^6 \,\GeV \cdot {\rm m}^{-3}$
(corresponding to one proton per cubic centimeter) \cite{WdB}. Extrapolating this 
number to Planck mass particles we would get $\sim 10^{-13}$ gravitinos per cubic 
meter, hence an extremely dilute gas of DM particles. Most of these will be color singlet gravitinos, 
whereas the strongly interacting non-singlets make up only a small fraction of these, see below.
Now in \cite{MN1}, we have already raised the possibility that DM, while more or less 
uniformly distributed  in interstellar space, might be subject to larger {\em local} variations near 
stars. This could happen if the DM co-rotates with the stars around the center of the galaxy, but 
not relative to them, unlike the dust that gives rise to planets and ends up rotating around 
(and not being absorbed by) the star. Then, a typical star could eat up much of  the 
surrounding DM in its vicinity over its lifetime, depleting a ball of diameter of a few
lightyears around it. Taking half of the average distance between two stars as a 
rough guide this would yield a volume of (two lightyears)$^3 \sim 10^{49} \, {\rm m}^3$ 
which in turn would imply a total amount of DM {\em within} the star of $\lesssim 10^{27}$ kg, 
corresponding to a fraction of $\lesssim 10^{-3}$ of its total mass for a pre-supernova star.
Although seemingly non-negligible it corresponds to only one gravitino per 10$^{22}$ protons 
or helium nuclei. This number is far too small to affect standard stellar processes in any 
significant way, especially since our gravitinos cannot decay into SM particles.
As participants in strong and electromagnetic interactions, they would nevertheless be 
in thermal equilibrium, with a very small velocity dispersion of $(\Delta v)^2 \sim kT/\MPL$. 
In contradistinction to luminous matter (for which $\Delta v$ is very large), they would 
thus continue to slowly migrate towards the center of the star, especially if the latter 
develops a core of heavier nuclei, thereby avoiding the usual problem of 
`missing the center' \cite {BW}.

At this stage the gravitino density is still so small that annihilation processes can be
neglected. Furthermore, because the color singlet gravitinos in (\ref{GravCharges}) 
interact only electromagnetically their annihilation cross section is proportional to 
the inverse mass squared ($\si v\sim (\pi\al^2\hbar^2/(4 M^{2}c)$ for small initial velocities),
hence of no significance to the effects considered here.
The situation 
is expected  to be entirely different for the strongly interacting gravitinos (or rather the 
associated color singlet bound states) corresponding to the two color triplets in 
(\ref{GravCharges}), so let us explain in more qualitative terms why this is so. 
Like standard QCD bound states these composite particles would come with
a strongly interacting cloud of gluons, and it is this fact that enhances the cross section. This
can be seen by inspection of the annihilation cross section for heavy strongly interacting 
particles. From unitarity arguments one can derive the following formula for the 
inelastic cross section $\si_{\rm inel}$ \cite{Lam}, which for particle anti-particle collisions
at sufficiently small kinetic energies is the same as the annihilation cross section,
\beq
\si_{\rm inel}  \,=\,  \frac{\pi}{k^2(2s_1+1)(2s_2+1)}\sum_j\sum_{l,s,n}(2j+1) X_{j \ell s n}
\label{crossbound}
\eeq
where \cite{Lam} 
\beq
X_{j\ell s n} \,=\,   1 - \big| \langle \ell s n| \rS |\ell s n\rangle \big|^2 -
  \sum_{\ell'\neq\ell, s'\neq s} \big|\langle \ell' s' n|(\E -\rS)|\ell s n\rangle \big|^2 
\eeq
where $\rS$ denotes the $S$-matrix ($j,\ell,s$ are the usual angular momentum labels,
while $n$ stands for any other quantum numbers).
The crucial point is now that for pointlike ({\em e.g.} electromagnetic) interactions, 
only the lowest angular momenta are relevant, so that with  ${\bf k}^2=M^2 v^2/4$ we get 
the expected result $\si_{\rm inel}v\sim 1/(M^2 v)$. By contrast, for strongly interacting 
particles many more angular momentum states may contribute to the sum in (\ref{crossbound}):
the known cross section for proton anti-proton collisions suggests that angular momenta 
get excited up to some maximum $j_{\rm max}\sim M/\La_{\rm QCD}$. Then the
sum over angular momenta  can compensate the smallness of the factor $1/M^2$ 
to replace it by $1/\La_{\rm QCD}^2$, whence we obtain
\beq
\si_{\rm inel}v \,\sim \, \frac{1}{\La_{QCD}^2}
\eeq
One can then fit the known proton-antiproton annihilation cross section to get the 
approximate formula \cite{Wibig} 
\bea\label{siv}
\langle \si \beta \rangle &\!\sim\!& \La_{QCD}^{-2} \left[ 3.84 - 0.51\ln\left(\frac{\sqrt{s}}{\La_{QCD}}\right)+
0.084 \left(\ln\left(\frac{\sqrt{s}}{\La_{QCD}}\right)\right)^2\right] 
\eea
with $\La_{QCD} = 0.2$ GeV. For $\sqrt{s}=\La_{QCD}$ we obtain $\langle \si \beta \rangle \sim 38\,$ mb.
This formula is non-perturbative in the sense that it does not rely on a perturbative 
calculation, but rather on imposing the Froissart bound  and fitting the relevant 
parameters to the data over the whole range of available energies (notice that 
the dependence on $s$ is not very pronounced). Putting $\sqrt{s} = 2\gamma m_p$
and $\gamma\sim 1$ we find  $\langle \si\beta\rangle \sim 32 {\rm mb}$ 
(1 mb $\sim 10^{-31} {\rm m}^2 \sim 2.5 \, \GeV^{-2}$). Again reasoning by analogy, we will 
use this value also for gravitino antigravitino annihilation to get an order of 
magnitude estimate.

To estimate the present density $\rho_0$ of strongly interacting 
(color triplet) gravitinos, we observe  that with a gravitino mass close to $\MPL$, 
the usual requirement of thermal equilibrium reads
\beq
\Ga=\rho \langle \si v \rangle \,>\,  H = \frac{\pi(k_BT)^2}{3\sqrt{5}\hbar c^2 \MPL}
\eeq
Adopting from now on the usual unit conventions $\hbar =c = k_B =1$
(hence $2 \cdot 10^{-7} \eV \cdot {\rm m} = 1$), and using the values for $\si$ given above
this translates into an equation for the relic abundance $\rho_T$
\beq
(32 \,{\rm mb}) \, \rho_T \equiv
(32 \, {\rm mb}) \,  g\left(\frac{mT}{2\pi}\right)^{3/2}  \re^{-m/T} =\frac{T^2}{2\MPL}
\eeq
($g=4$ for a massive gravitino), or
\beq
\frac{m}{T}\sim 90\ \ \; \Ra\ \ \rho_T\sim  3\cdot 10^{59} \, {\rm m}^{-3}
\eeq
The temperature $T \sim 2\cdot10^{16}$ GeV corresponds to cosmic time $t_T = \MPL/T^2
\sim 3\cdot 10^{-39}$ s. The {\em present} density $\rho_0$ is obtained from
$\rho_T$ by the well known formula
\beq
\rho_0=\rho_T\left(\frac{a_T}{a_0}\right)^3
\eeq
(since superheavy gravitinos are non-relativistic).
Taking the end of the radiation dominated era as $10^{12}\,$s we get
\beq
\frac{a_T}{a_0} =\left(\frac{3\cdot 10^{-39}}{10^{12}}\right)^{1/2}
\left(\frac{10^{12}}{3\cdot 10^{17}}\right)^{2/3}   \,\sim \, 10^{-29}
\eeq
where the two factors correspond to the radiation dominated and matter dominated
eras, respectively. Thus
\beq
\rho_0\,\sim \, 5\cdot 10^{-28}\ {\rm m}^{-3}\ \ \ \ \big(\sim 10^{-9}\ {\rm GeV}\cdot{\rm m}^{-3}\big)
\eeq
Assuming now (as before) that the star `swallows' all gravitinos within a radius of
two lightyears we get for the total number of color triplet gravitinos inside the star
\beq
N_g \,\sim \, 2\cdot 10^{22}
\label{NSun}
\eeq
As we argued above, the gravitinos inside the star are in thermal, but not mechanical 
equilibrium, so for a pre-supernova star we expect them to cluster more towards the 
iron core (where no nuclear reactions take place anymore).
As already pointed out, the number (\ref{NSun}) is too small to produce any significant
effects in the star -- even in the iron core the lifetime of gravitinos via annihilation still exceeds the 
lifetime of the Universe, see below. 

The situation changes dramatically if the star collapses to a neutron star. 
In that case, as explained above, most of the gravitinos will be contained in its iron core
even prior to the supernova collapse, and the gravitinos will collapse with 
the core due to the sudden increase of gravititional pull towards the center. 
As a consequence, they will get squeezed into a 
ball of radius $\cO(10 \, {\rm km})$ \cite{ST}, and their density increases to 
\beq\label{ng}
\rho_{NS}\, \sim \, 5\cdot10^{9} \, {\rm m}^{-3}. 
\eeq
This `compactification' is absolutely crucial since the gravitinos
need to be packed sufficiently closely to enable them to annihilate in any
appreciable rate.  The inverse lifetime of the gravitino as a function of the neutron star time from its birth is
\beq
\Ga_{NS} (t)=\rho_{NS} \exp\left(-\int_0^t \Ga_{NS} (t')\rd t'\right)\langle \si v \rangle\; 
\eeq
which gives 
\beq
\Ga_{NS} (t)=\frac{\Ga_{NS}(0)}{1 + \Ga_{NS}(0) t}
\eeq
with the initial value (and $\langle \si\beta\rangle \sim 32 \, {\rm mb}$)
\beq\label{GaNS1}
\Ga_{NS} (0)\sim \big(5\cdot 10^{9}\big)\cdot (32\cdot 10^{-31}) \cdot \big(3\cdot 10^8\big) {\rm s}^{-1}
\,\sim\,  5\cdot 10^{-12} {\rm s}^{-1}  
\eeq
Therefore the actual annihilation rate depends on the age of the neutron star. We also see 
that before the collapse for a pre-supernova star with an iron core of $\cO(1000\,{\rm km})$
diameter the rate would be lower by a factor of $\sim 10^{-12}$.

Having derived the approximate annihilation rate inside the neutron stars we can 
now estimate the number of UHECR particles coming from the annihilation. 
Because the superheavy gravitinos interact strongly, each single annihilation will result 
in a violent burst of Planck scale energy, producing a multitude of (mostly hadronic) 
particles. We can roughly estimate their multiplicity by extrapolating to Planckian 
energies the formula \cite{Mult}
\beq\label{mult}
{\rm multiplicity} \,\sim\, 0.27 \,\al_s(\Lambda)\exp\left(\frac{2.26}{\sqrt{\al_s(\Lambda)}}\right) 
 \eeq
whose validity has been confirmed for proton-proton collisions from the lowest energies to 
the highest energies attainable by LHC; this must suffice for the present purposes as no 
other formula seems to be available for the much larger energies considered here.
The strong coupling $\alpha_s$ is to be evaluated at $\Lambda \sim 0.35 \sqrt{s}$.
Plugging in $\sqrt{s}\sim \MPL$ gives $\cO(10^6)$ particles per annihilation (whereas for 
the cross section in (\ref{siv}) the relevant quantity is the relative velocity $\gamma$ 
which in our case is very close to one). The total  energy 
$\sim \MPL$ will be distributed over all these particles, with an average energy of 
$10^{13} \GeV \sim 10^{22} \eV$ per  particle. This  happens to be of the same 
order of  magnitude as the maximum energy observed in high energy cosmic rays! 
Nevertheless, because of their strong interactions and the large density inside the 
neutron star the annihilation products cannot escape because they will either lose 
too much energy or be stopped altogether on their way out from the core of the neutron star. 
For this reason, we expect the main contribution to UHECR particles to come from 
the outermost shell of the neutron star of width $d \lesssim \cO(100 \,{\rm m})$ {\em i.e.}
$\sim 3\%$ of the volume. There the density drops down by a factor $10^{-5}$ relative to 
the core density  \cite{ST}, and is given by $\sim 10^{13}$ kg m$^{-3}$ such that
 $\rho(R-d)=10^{40}$ m$^{-3}$. (The  width $d$ is here determined by the requirement 
that the number of collisions times the loss of energy per collision, here assumed 
to be $\sim \cO(1\!\GeV)$, should be much lower than the total energy of the proton 
of $10^{12}$ GeV, which gives $\rho(R-d) \si d< 10^{12}$, where $\si \sim 10^{-30}$ m$^2$).

Importantly, this outer shell is thought to be rich in heavier nuclei, including iron nuclei \cite{ST}, 
so the high energy particles that can escape may `sweep up' hadrons as well as
heavier ions before exiting the neutron star.  Whether and how such a `sweep-up' 
could work in detail remains to be explored, as this would require a much better 
understanding of difficult issues in neutron star and strong interaction physics.
Moreover, no data that could be relevant to such processes appear to be available in 
the published literature (for instance, the ALICE experiment at CERN does not provide 
any information on heavy fragments with large longitudinal momentum resulting from 
the proton-lead collisions). If indeed a sufficient abundance of such highly energetic fragments 
could be generated and get out of the neutron star, the resulting heavier ions 
will subsequently decay to a variety of stable isotopes, and 
thus end up as ultra-high energy stable ions of the type observed by \cite{AUGER}.

Despite the fall-off of the density profile towards the outer regions of the neutron star
the density of gravitinos near the skin may actually be enhanced by a `centrifuge effect' 
for rapidly spinning neutron stars. These two effects would have to compete with 
each other in determining the gravitino density,
but due to the lack of sufficiently detailed information on the physics of neutron stars
not much more can be said at this point. For this reason we shall simply take 
the value (\ref{ng}) to hold also near the skin of the neutron star, and assume that $3\%$ 
of the neutron star volume is effectively available for this process. A young neutron star 
would thus continuously `spray' high energy protons or heavy ions at a rate
\beq
\sim  6\cdot 0.03\cdot (2\cdot 10^{22})\cdot  (5\cdot 10^{-12}) \cdot 10^6 {\rm s}^{-1}
\sim 2\cdot 10^{16} {\rm s}^{-1}
\eeq
from its surface into outer space (the factor of 6 comes from the number of the 
strongly interacting gravitino species). 
To calculate how many of these will eventually reach Earth, we recall that, with an 
estimated average number of neutron stars per galaxy of $\sim 10^8$ and $\sim 10^{7}$ 
galaxies within a GKZ horizon  of 50 Mpc \cite{UHE2}, we have a total number of 
$10^{15}$  UHECR emitters. Denoting the density of neutron stars in the universe by $\rho_N(\bx)$ (where 
$\bx=0$ corresponds to the position of Earth), the total rate arriving at Earth is thus
\beq\label{cN1}
N_E \,\sim \, \big( 2\cdot 10^{16} s^{-1}\big)  \times \int \frac{\rho_N(\bx) d^3x}{4\pi |\bx|^2}
\eeq
For a rough estimate of the total flux we neglect density variations, taking
$\rho_N =$ const, in which case the integral is easily evaluated to be
\beq
N_E \,\sim \, \rho_N R_{max}   \times 2\cdot 10^{16} {\rm s}^{-1} 
\eeq
Putting $R_{max}\sim 50 \!$ Mpc as a cutoff we arrive at the flux of UHECR 
arriving on Earth as
\beq
N_E\sim\frac{10^{15}\cdot 2 \cdot 10^{16}}{4( 10^{24})^2}\ {\rm m}^{-2} {\rm s}^{-1}
\,\sim\,  5\cdot 10^{-18} \ {\rm m}^{-2} {\rm s}^{-1}
\eeq
which is not too far off the observed rate of one UHECR event per 
month and per 3000 km$^2$ \cite{AUGER}. To be sure, the UHECR emitters
are not evenly distributed throughout the universe, and we therefore expect an increased 
number of events to originate from superclusters of galaxies  rich in neutron stars
(the supergalactic plane, in particular, as also suggested by the data \cite{AUGER}). 
In particular the integral in
(\ref{cN1}) may receive its dominant contribution from a disk rather than the full ball.
We also note that with a maximum available energy of $\cO(10^{22} \eV)$
our proposal can also explain the existence of (very rare) UHECR events {\em exceeding} 
the GKZ bound, if these originate from neutron stars {\em within} the Milky Way or nearby galaxies.

We thus arrive at an explanation which agrees qualitatively with observations, and at an 
estimated event rate that albeit subject to some crucial assumptions,
is not too far from the one observed. Evidently,  there remain
many uncertainties in our calculation, quite apart from questions concerning the viability 
of the unification scenario proposed in \cite{MN0}. 
Nevertheless, we find it  remarkable that the present proposal could tie in with the scheme 
proposed in \cite{MN0} to explain the fermion content of the SM, with three generations 
of quarks and leptons.

\vspace{1mm} 
\noindent
 {\bf Acknowledgments:} 
We are most grateful to Masaru Shibata and Kacper Zalewski for enlightening discussions. 
K.A.~Meissner thanks AEI for hospitality and support; he was 
 partially supported by the Polish National Science Center grant DEC-2017/25/B/ST2/00165.
 The work of  H.~Nicolai has received funding from the European Research 
 Council (ERC) under the  European Union's Horizon 2020 research and 
 innovation programme (grant agreement No 740209).

\end{document}